\documentclass[
aps,%
12pt,%
final,%
notitlepage,%
oneside,%
onecolumn,%
nobibnotes,%
nofootinbib,% 
superscriptaddress,%
noshowpacs,%
centertags]%
{revtex4}

\usepackage{bm}
\usepackage{graphics}
\usepackage{rotating}
\usepackage{amsmath}
\usepackage{amssymb}
\usepackage{epsfig}

\begin{document}

\title{Production of heavy tetraquarks in rare exclusive decays\\ of the Higgs boson}
\author{\firstname{F.~A.} \surname{Martynenko}\footnote{f.a.martynenko@gmail.com}}
\affiliation{Samara University, Samara, Russia}
\author{\firstname{A.~V.} \surname{Eskin}}
\affiliation{Samara University, Samara, Russia}
\author{\firstname{A.~P.} \surname{Martynenko}}
\affiliation{Samara University, Samara, Russia}

\begin{abstract}
The relativistic quark model is used to study the processes of heavy tetraquark $(cc\bar c\bar c)$ 
production in the Higgs boson decay. 
We consider quark-gluon, ZZ-boson, and quark-gluon loop decay mechanisms with the production 
of a fully charmed tetraquark.
Relativistic interaction amplitudes are constructed and decay widths 
are calculated taking into account relativistic corrections in the decay amplitude connected 
with the relative motion of quarks and antiquarks.
\end{abstract}

\maketitle

\section{Introduction}

The investigation of production processes of heavy quark bound states in various nuclear reactions 
has been and remains an important area of work. Such reactions allow to study the properties 
of resonances being produced, the interaction constants of particles of the Standard Model, 
the bound states production and decay mechanisms, and to test the models used to describe bound states of quarks. 
Such processes are studied both theoretically and experimentally. It can be noted that very often 
at the initial stage, theoretical work is ahead of experimental work. For example, long before 
the discovery of the Higgs boson, the reactions of its production and decay were studied theoretically. 
Such work served as a kind of guideline for experimenters in setting up experiments. But as soon 
as experimental data appeared, 
it was usually necessary to refine the theoretical calculations and implement them with greater accuracy.

In the last decade, exotic bound states of heavy quarks that do not fit the standard quark model, 
most notably tetraquarks and pentaquarks, have come under intense study. The discovery 
of the charmonium-like state X(3872) by the Belle collaboration in 2003 brought the study 
of exotic states to the forefront of hadron physics. Although many tetraquark candidates containing 
heavy (c,b) quarks are now known \cite{t1,t2,cms1,t3,t3a,t3b,t4,t4a,t5,t6,t7}, 
many questions remain about them, including which 
of them are truly exotic hadrons and what their internal structure is.

For the production of exotic bound states, one can use well-developed methods that have been successfully 
applied to the production of ordinary heavy mesons and baryons. In this paper, we study the processes 
of heavy tetraquark production in the Higgs boson decay. We further consider the production of a vector 
tetraquark consisting of two $c$-quarks and two $\bar c$-antiquarks in the decay $H \to T_{cc\bar c\bar c}+\gamma$. 
The LHCb collaboration detected the X(6900) resonance, which was confirmed 
in the ATLAS and CMS experiments
and is a candidate for the tetraquark $(cc\bar c\bar c)$ along with numerous other 
resonances discovered in recent years \cite{t1,t2,cms1,t3,t3a,t3b,t4,t4a,t5,t6,t7}. 
An analysis of the invariant mass distribution in the proton-proton 
interaction \cite{cms1} revealed three resonance structures: two new states $X(6600)$ and $X(7100)$, 
and confirmation of the previously observed $X(6900)$. The first determination of quantum numbers 
for fully charmed tetraquarks was performed.
In connection with the 
appearance of experimental data on the production of exotic bound states (tetraquarks) of quarks and gluons, 
a significant number of papers have appeared in which the mass spectrum and cross sections of their production 
in proton-proton interactions, photoproduction, electron-positron annihilation, and B-meson decays 
were calculated \cite{t7,brodsky1,t8,t9,t9a,t9b,t9c,t9d,t9e,t10,t11,t12,t13,t14,tc1,tc2,tc3} (see other references 
in review papers \cite{t4,t4a,t5}). 

The production of tetraquarks in the decay of the Higgs boson $H\to T_{cc\bar c\bar c}+\gamma $ 
is another channel for their study. Despite the fact that such processes are rare, their study is 
of interest in connection with the construction of various Higgs boson factories. The goal of this paper 
is to obtain estimates for corresponding Higgs boson decay widths. In doing so, we use the relativistic 
approach to studying the reactions of quark bound state production, which was previously applied 
for various processes \cite{apm2023,apm2022,apm2024}. 
In the case of pair production of charmonium in the decay of the $H$ boson, which we studied earlier in 
\cite{apm2023}, two bound states $(c\bar c)$ arise in the final state. In this regard, it was interesting 
to study the production of a bound state of four particles $(cc\bar c\bar c)$ (tetraquark 
$T_{cc\bar c\bar c}$) also in the decay of the Higgs boson in the relativistic quark model. 
We concentrate on the study of the production of a tetraquark built from quarks and antiquarks 
of the same kind, using a configuration of two pairs of particles $[cc][\bar c\bar c]$ for its construction,
where square brackets denote the color antitriplet state. 
Such model is close in structure to the diquark-antidiquark model, although the tetraquark is 
generally treated as a four-particle state, the calculation of wave functions and energy levels of which 
can be carried out within the variational method. Within relativistic quark model, it is possible 
to consistently calculate relativistic effects due to relative motion of quarks forming a bound state. 
In the production of bound states of $c$-quarks, relativistic corrections are essential for achieving 
high accuracy of calculation.

\section{General formalism}

\begin{figure}[htbp]
\centering
\includegraphics[scale=1.2]{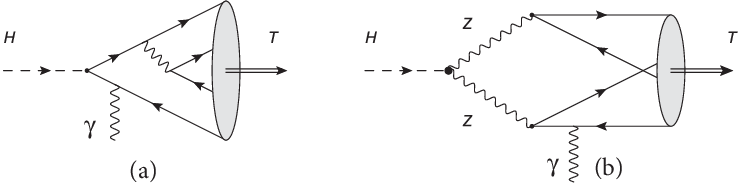}
\caption{Quark-gluon and $ZZ$-boson mechanisms of heavy tetraquark production in the Higgs boson 
decays $H\to T+\gamma$}.
\label{fig1}
\end{figure}

\begin{figure}[htbp]
\centering
\includegraphics[scale=1.2]{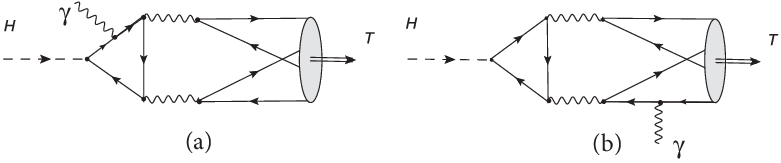}
\caption{Quark-gluon loop mechanism of heavy tetraquark production in the Higgs boson decay $H\to T+\gamma$}.
\label{fig2}
\end{figure}

There are various mechanisms of tetraquark production in the decay $H\to T_{cc\bar c\bar c}+\gamma $: 
quark-gluon (Fig.~\ref{fig1}(a)), ZZ-boson (Fig.~\ref{fig1}(b)), quark-gluon (photon) loop (Fig.~\ref{fig2}(a,b)), 
W-boson loop and others. The differences between them are connected with physical processes that occur 
at the microscopic level and ultimately lead to the formation of a tetraquark in the final state. 
The amplitudes of such production mechanisms differ in the degrees of interaction constants $\alpha$ 
(fine structure constant), $\alpha_s$ (strong interaction constant), as well as in characteristic 
ratios of particle masses.

In the case of the $ZZ$-boson mechanism, there are two vertices of particle interaction. 
One vertex of the $H\to ZZ$ interaction is determined by the factor \cite{CORE}:
\begin{equation}
\label{f1}
V^{\alpha\beta}=\frac{2e}{\sin 2\theta_W} M_Z g^{\alpha\beta}.
\end{equation}

The second vertex of interaction of each $Z$-boson with $c$-quarks has the form \cite{CORE}:
\begin{equation}
\label{f2}
\Gamma^{\alpha}=\frac{e}{\sin 2\theta_W} \gamma^\alpha
\left[\frac{1}{2}(1-\gamma_5)-a_z\right],~~~a_z=2Q_c\sin^2\theta_W.
\end{equation}

For the quark-gluon mechanism, the interaction vertex $H\to c\bar c$ includes the factor:
\begin{equation}
\label{f3}
V(H\to c\bar c)=-\frac{e}{\sin 2\theta_W} \frac{m}{M_Z},
\end{equation}
where $m$ is the mass of $c$-quark, $M_Z$ is the mass of $Z$ boson.

Let us consider the $ZZ$-boson mechanism of tetraquark production shown in Fig.~\ref{fig1}(b) in the reaction 
$H\to T_{cc\bar c\bar c}+\gamma$. Our goal is to obtain an estimate of the decay width of the Higgs 
boson with the production of the tetraquark $T_{cc\bar c\bar c}$ taking into account relativistic 
corrections, determined by the momenta of relative motion of heavy quarks during the formation of a tetraquark.

The color structure of the tetraquark state $(cc\bar c\bar c)$ can be either of the hadronic molecule type:
$3_c\otimes 3_c\otimes \bar 3_{\bar c}\otimes \bar 3_{\bar c}\to 1_{[\bar c c]_1[\bar c c]_1}$, or of the 
diquark-antidiquark type
$3_c\otimes 3_c\otimes \bar 3_{\bar c}\otimes \bar 3_{\bar c}\to 1_{[ c c]_{\bar 3}[\bar c \bar c]_3}\oplus 1_{[ c c]_{6}[\bar c \bar c]_{\bar 6}}$. Pairs of quarks in the color antitriplet state are attracted to each other, 
and in the color sextet state they repel each other. Similarly, in a quark-antiquark pair, the attractive 
force arises in the color singlet state, and in the color octet state, the repulsive force arises.
Thus, various configurations of four quarks are possible, which lead to the formation of a tetraquark \cite{t10,t12,tetra1,tetra2,tetra3,tetra4}. We consider the case of production, when a tetraquark 
is formed from two pairs quark-quark and antiquark-antiquark, each of which has a spin 1 and is 
in a color-antisymmetric state. When a tetraquark consists of identical quarks and antiquarks, the Pauli 
principle leads to additional restrictions on the quantum numbers of a pair of identical quarks (diquark). 
Indeed, permutation of quark indices should change the sign of total wave function of a pair of quarks. 
The color part of wave function of a pair of quarks is antisymmetric. The coordinate wave function 
is symmetric, since the quarks are in the S-wave state, therefore the spin part of wave function 
should also be symmetric. Therefore, total spin of the $S$-wave state of the quark pair must 
be equal to 1.
In terms of the configuration we are considering (diquark-antidiquark), total angular momentum ${\bf J}$ 
of fully charmed tetraquark $(cc\bar c\bar c)$ is obtained by adding total angular momentum ${\bf J}_{cc}$ 
of doubly charmed state $(cc)$, total angular momentum ${\bf J}_{\bar c\bar c}$ of doubly 
charmed state $(\bar c\bar c)$, and relative orbital angular momentum ${\bf L}_\sigma$ (relative 
coordinates ${\boldsymbol\rho}$, ${\boldsymbol\lambda}$, ${\boldsymbol\sigma}$ are 
defined in \cite{tetra2025}) between the state $(cc)$ and the state $(\bar c\bar c)$.


Introducing the total $P$, $Q$ and relative $p$, $q$ four-momenta
in pairs $(cc)$, $(\bar c\bar c)$, we can express the momenta of individual quarks and antiquarks 
as follows:
\begin{equation}
\label{f4}
p_1=\frac{1}{2}P+p,~~~p_2=\frac{1}{2}P-p,~~~q_1=\frac{1}{2}Q+q,~~~q_2=\frac{1}{2}Q-q.
\end{equation}

Let us further introduce total $T$ and relative $t$ four-momenta for these pairs:
\begin{equation}
\label{f5}
P=\frac{1}{2}T+t,~~~Q=\frac{1}{2}T-t.
\end{equation}

Denoting the photon momentum $k$, and the Higgs boson momentum $r$, we obtain from 
the law of momentum conservation:
\begin{equation}
\label{f6}
r=k+T,~~~kT=\frac{1}{2}(M_H^2-M_T^2),~~~rT=\frac{1}{2}(M_H^2+M_T^2),
\end{equation}
where $M_T$ is the tetraquark mass, $M_H$ is the Higgs boson mass.

Total angular momentum of a tetraquark, obtained by adding two spins 1 of pairs of quarks $(cc)$ 
and antiquarks $(\bar c\bar c)$ with zero orbital angular momentum of quarks, can take three values 
$J = 0, 1, 2$, so that the vectors of corresponding states can be represented as follows:
\begin{eqnarray}
\label{f6a}
|0^{++}>_T=|S_{cc}=1,S_{\bar c\bar c}=1,S_T=0,L_T=0>_{J_T=0},\\
\label{f6b}
|1^{+-}>_T=|S_{cc}=1,S_{\bar c\bar c}=1,S_T=1,L_T=0>_{J_T=1},\\
\label{f6c}
|2^{++}>_T=|S_{cc}=1,S_{\bar c\bar c}=1,S_T=2,L_T=0>_{J_T=0},
\end{eqnarray}
where the spatial and charge parities of the tetraquark are determined by the formulas:
\begin{equation}
\label{f6d}
P_T=(-1)^{L_T},~~~C_T=(-1)^{S_T+L_T}.
\end{equation}
The photon is emitted by one of quarks or antiquarks, so for the $ZZ$-boson production mechanism 
we have four amplitudes in Fig.~\ref{fig1}(b).

The original expression for relativistic decay amplitude is a convolution of the production amplitude 
of two quarks $c$, two antiquarks $\bar c$, and a photon with the Bethe-Salpeter amplitude 
describing the bound state of four particles. This convolution has the same general structure 
as in the case of two particles \cite{apm2006,symmetry}.
After transforming this convolution to a three-dimensional quasipotential form, we obtain 
a convolution of the interaction amplitude ${\cal T}({\bf p},{\bf q},{\bf t},T,k)$ projected 
onto positive-frequency states and the tetraquark quasipotential wave function 
$\Psi_T({\bf p},{\bf q},{\bf t})$:
\begin{equation}
\label{f7}
{\cal M}(H\to T+\gamma)=\int\frac{d{\bf p}}{(2\pi)^3}\int\frac{d{\bf q}}{(2\pi)^3}\int\frac{d{\bf t}}{(2\pi)^3}
\bar\Psi_T^{\lambda_1\lambda_2\lambda_3\lambda_4}({\bf p},{\bf q},{\bf t})
{\cal T}^{\lambda_1\lambda_2\lambda_3\lambda_4}({\bf p},{\bf q},{\bf t},T,k).
\end{equation}

The law of transformation of tetraquark wave function during a transition from a reference frame 
with momentum $T$ to a reference frame at rest is similar in structure to the case of a two-particle 
bound state \cite{brodsky,faustov}:
\begin{equation}
\label{f8}
\bar\Psi_T^{\mu\nu\lambda\sigma}({\bf p},{\bf q},{\bf t})=
\bar\Psi_0^{\alpha\beta\rho\omega}({\bf p},{\bf q},{\bf t})
D_1^{+1/2~\alpha\mu}(R^W_{L_T})D_2^{+1/2~\beta\nu}(R^W_{L_T})
D_3^{+1/2~\rho\lambda}(R^W_{L_T})D_4^{+1/2~\omega\sigma}(R^W_{L_T}),
\end{equation}
where $R^W$ is the Wigner rotation, $L_{T}$ is the Lorentz boost from the tetraquark rest frame 
to the moving reference frame, $D^{1/2}(R)$ is the rotation matrix, which is defined as follows:
\begin{equation}
\label{f9}
{1 \ \ \,0\choose 0 \ \ \,1}D^{1/2}_{1,2}(R^W_{L_{T}})=
S^{-1}({\bf p}_{1,2})S({\bf T})S({\bf p}),~~~
{1 \ \ \,0\choose 0 \ \ \,1}D^{1/2}_{3,4}(R^W_{L_{T}})=
S^{-1}({\bf q}_{1,2})S({\bf T})S({\bf q}),
\end{equation}
where the matrix of the Lorentz transformation of the Dirac bispinor has the form:
\begin{equation}
\label{f10}
S({\bf p})=\sqrt{\frac{\epsilon(p)+m}{2m}}\left(1+\frac{(\bm{\alpha}
{\bf p})} {\epsilon(p)+m}\right).
\end{equation}

To further transform the amplitude \eqref{f7} we use the following formulas for transforming 
the Dirac wave functions \cite{faustov}:
\begin{eqnarray}
\label{f11a}
S_{\alpha\beta}(\Lambda)u^\lambda_\beta(p)=\sum_{\sigma=\pm 1/2}
u^{\sigma}_\alpha(\Lambda p)D^{1/2}_{\sigma\lambda}(R^W_{\Lambda p}),\\
\label{f12a}
\bar u^\lambda_\beta(p)S^{-1}_{\beta\alpha}(\Lambda)=\sum_{\sigma=\pm 1/2}
D^{+~1/2}_{\lambda\sigma}(R^W_{\Lambda p})\bar u^\sigma_\alpha(\Lambda p).
\end{eqnarray}

As a result of transformations using \eqref{f8}-\eqref{f12a}, the expressions for four 
decay amplitudes in the case of the $ZZ$-boson mechanism take the form:
\begin{equation}
\label{f11}
{\cal M}_{T+\gamma}^{(1)}=\frac{e^4}{4\sin 2\theta M_Z m^2}\int d{\bf p}\int d{\bf q}\int d{\bf t}
\bar\Psi_0({\bf p,\bf q,\bf t})D^{\alpha\lambda}_Z(k_1)D^{\alpha\sigma}_Z(k_2)
\frac{N({\bf p,\bf t})N({\bf q,\bf t})}{\bigl[\frac{1}{4}\bigl(M_H^2-
\frac{3}{4}M_T^2\bigr)-m^2\bigr]}\times
\end{equation}
\begin{displaymath}
Tr \Bigl\{
\hat\Pi_1^V({\bf p,\bf t})\hat\varepsilon_\gamma(\hat p_1+\hat k+m)\Gamma^\lambda
\hat\Pi_2^V({\bf q,\bf t})\Gamma^\sigma \Bigr\},
\end{displaymath}
\begin{equation}
\label{f12}
{\cal M}_{T+\gamma}^{(2)}=\frac{e^4}{4\sin 2\theta M_Z m^2}\int d{\bf p}\int d{\bf q}\int d{\bf t}
\bar\Psi_0({\bf p,\bf q,\bf t})D^{\alpha\lambda}_Z(k_1)D^{\alpha\sigma}_Z(k_2)
\frac{N({\bf p,\bf t})N({\bf q,\bf t})}{\bigl[\frac{1}{4}\bigl(M_H^2-
\frac{3}{4}M_T^2\bigr)-m^2\bigr]}\times
\end{equation}
\begin{displaymath}
Tr \Bigl\{
\hat\Pi_1^V({\bf p,\bf t})\Gamma^\lambda
\hat\Pi_2^V({\bf q,\bf t})\Gamma^\sigma (\hat p_1+\hat k+m)\hat\varepsilon_\gamma \Bigr\},
\end{displaymath}
\begin{equation}
\label{f13}
{\cal M}_{T+\gamma}^{(3)}=\frac{e^4}{4\sin 2\theta M_Z m^2}\int d{\bf p}\int d{\bf q}\int d{\bf t}
\bar\Psi_0({\bf p,\bf q,\bf t})D^{\alpha\lambda}_Z(k_1)D^{\alpha\sigma}_Z(k_2)
\frac{N({\bf p,\bf t})N({\bf q,\bf t})}{\bigl[\frac{1}{4}\bigl(M_H^2-
\frac{3}{4}M_T^2\bigr)-m^2\bigr]}\times
\end{equation}
\begin{displaymath}
Tr \Bigl\{
\hat\Pi_1^V({\bf p,\bf t})\Gamma^\lambda (-\hat q_1-\hat k+m)\hat\varepsilon_\gamma
\hat\Pi_2^V({\bf q,\bf t})\Gamma^\sigma  \Bigr\},
\end{displaymath}
\begin{equation}
\label{f14}
{\cal M}_{T+\gamma}^{(4)}=\frac{e^4}{4\sin 2\theta M_Z m^2}\int d{\bf p}\int d{\bf q}\int d{\bf t}
\bar\Psi_0({\bf p,\bf q,\bf t})D^{\alpha\lambda}_Z(k_1)D^{\alpha\sigma}_Z(k_2)
\frac{N({\bf p,\bf t})N({\bf q,\bf t})}{\bigl[\frac{1}{4}\bigl(M_H^2-
\frac{3}{4}M_T^2\bigr)-m^2\bigr]}\times
\end{equation}
\begin{displaymath}
Tr \Bigl\{
\hat\Pi_1^V({\bf p,\bf t})\Gamma^\lambda 
\hat\Pi_2^V({\bf q,\bf t})\hat\varepsilon_\gamma(-\hat q_2-\hat k+m)
\Gamma^\sigma  \Bigr\},
\end{displaymath}
where the relativistic normalization factor has the form:
\begin{equation}
\label{f15}
N({\bf p,\bf t})=\left[
\frac{\epsilon({\bf p}+{\bf t/2})}{m}\frac{(\epsilon({\bf p}+{\bf t/2})+m)}{2m}
\frac{\epsilon({\bf p}-{\bf t/2})}{m}\frac{(\epsilon({\bf p}-{\bf t/2})+m)}{2m}
\right]^{-1/2}.
\end{equation}

The momenta in the $Z$-boson propagators are:
\begin{equation}
\label{f16}
k_1=\frac{1}{2}T+p+q+t,~~~k_2=\frac{1}{2}T-p-q.
\end{equation}

The amplitudes \eqref{f11}-\eqref{f14} contain relativistic projection operators onto the state 
of a pair of quarks or a pair of antiquarks with spin 1:
\begin{equation}
\label{f17}
\hat\Pi_i^V({\bf p},{\bf t})=
\bigl[\frac{\hat v-1}{2}-\hat v\frac{(\epsilon({\bf p}+\frac{\bf t}{2})-m)}{2m}-
\frac{(\hat p+\frac{\hat t}{2})}{2m}\bigr]\hat{\varepsilon_i}(1+\hat v) 
\bigl[\frac{\hat v+1}{2}-\hat v\frac{(\epsilon({\bf p}-\frac{\bf t}{2})-m)}{2m}+
\frac{(\hat p-\frac{\hat t}{2})}{2m}\bigr],
\end{equation}
where $\varepsilon_i$ (i=1,2) are polarization vectors describing pairs of quarks and antiquarks 
with spin 1. The color part of amplitudes \eqref{f11}-\eqref{f14} is allocated
as a product of color factors describing pairs of quarks and antiquarks in an antisymmetric 
state:
$\frac{1}{\sqrt{2}}\varepsilon_{ijk}\frac{1}{\sqrt{2}}\varepsilon_{ijl}=\delta_{kl}$.

Leaving in the numerator of amplitudes the quadratic terms in relative momenta 
$\omega_p={\bf p}^2/m^2$,
$\omega_q={\bf q}^2/m^2$, $\omega_t={\bf t}^2/m^2$, we obtain the following expression 
for the numerator of total amplitude of the production of a vector tetraquark and a photon:
\begin{equation}
\label{f18}
{\cal N}_{T+\gamma}=N_1(v\varepsilon_\gamma)(v_\gamma\varepsilon_T)+
N_2(\varepsilon_\gamma\varepsilon_T)+N_3\varepsilon_{\mu\nu\sigma\lambda}
v^\mu v_\gamma^\nu \varepsilon_T^\sigma \varepsilon_\gamma^\lambda,
\end{equation}
\begin{equation}
\label{f19}
N_1=-4(1-2a_z)\bigl(1-\frac{7}{48}r_2^2+\frac{7}{48}r_3^2-\frac{1}{192}r_2^2 r_3^2\bigr)
\omega_t,
\end{equation}
\begin{equation}
\label{f20}
N_2=-4(1-2a_z)\omega_t\left[r_3^{-1}\left(-\frac{1}{2}\frac{r_2^2}{r_3}-
\frac{1}{96}\frac{r_2^4}{r_3}\right)+\frac{1}{4}-
\frac{1}{24}r_2^2+\frac{1}{128}r_2^4+\frac{1}{32}r_3^2-\frac{1}{192}r_3^2r_2^2\right],
\end{equation}
\begin{equation}
\label{f21}
N_3=32-2r_2^2-4r_3^2+\frac{1}{2}r_2^2r_3^2-a_z(1-a_z)-(\omega_p+\omega_q)\Bigl[
-\frac{112}{3}+\frac{7}{3}r_2^2+\frac{14}{3}r_3^2-\frac{7}{12}r_2^2r_3^2+
\end{equation}
\begin{displaymath}
a_z(1-a_z)\bigl(\frac{224}{3}+\frac{2}{3}r_2^2+\frac{4}{3}r_3^2-\frac{1}{6}r_2^2r_3^2\bigr)\Bigr]-
\omega_t\Bigl[-\frac{61}{3}+\frac{59}{48}r_2^2+\frac{67}{24}r_3^2-\frac{65}{192}r_2^2 r_3^2+
\end{displaymath}
\begin{displaymath}
a_z(1-a_z)\bigl(\frac{122}{3}-\frac{1}{2}r_2^2-r_3^2+\frac{1}{8}r_2^2 r_3^2\bigr)
\Bigr],
\end{displaymath}
where auxiliary four-vectors $v=\frac{T}{M_T}$, $v_\gamma=\frac{k}{M_T}$
are introduced for convenience.
Parameters $r_1$, $r_2$, $r_3$ are determined by the ratios of particle masses in the form:
\begin{equation}
\label{f22}
r_1=\frac{m}{M_T},~~~r_2=\frac{M_H}{M_Z},~~~r_3=\frac{M_T}{M_Z}.
\end{equation}

The corresponding decay width, taking into account the leading order relativistic corrections, is:
\begin{equation}
\label{f30}
\Gamma(H\to ZZ\to \gamma+T)=
\frac{256\pi^3\alpha^4 r_3^2(r_2^2-r_3^2)|\Psi(0,0,0)|^2}{\sin^2 2\theta_W M_Z^8 r_2^3(4-r_3^2)^2
(2r_2^2-r_3^2-4)^2[\frac{1}{4}(r_2^2-\frac{3}{4}r_3^2)-r_1^2r_3^2]^2}\times
\end{equation}
\begin{displaymath}
\left[3N_2^2+2N_3^2(vv_\gamma)^2-N_1^2(vv_\gamma)^2\right],~~~(vv_\gamma)=
\frac{(r_2^2-r_3^2)}{2r_3^2}.
\end{displaymath}

One of parameters that enters \eqref{f30} is the value of the tetraquark wave function at zero 
coordinate values $\Psi(0,0,0)$. This parameter can be found by solving the eigenvalue problem for 
the Hamiltonian of four-particle system. As discussed above, total tetraquark wave function 
can be represented as the product of spin wave function $ \chi_s$, color wave function 
$\chi_c$, and coordinate wave function $\Psi$ in the general form:
\begin{equation}
\label{f30a}
\Phi={\cal A}\left(\chi_s\otimes\chi_c\otimes\Psi\right),
\end{equation}
where ${\cal A}$ is the antisymmetrization operator of two identical quarks and antiquarks. In the rest frame 
of the tetraquark, its coordinate wave function is obtained within the variational method as an expansion 
in the Gaussian trial functions. 
%For a system of four particles, there are three configurations for describing 
%the tetraquark $(cc\bar c\bar c)$. The first two have a dimeson configuration, and the third has 
%a diquark-antidiquark configuration, in which pairs of identical quarks and antiquarks form a color 
%triplet or antitriplet. It is this configuration that we consider in this paper. 
The quark interaction 
potential in nonrelativistic approximation can be represented as the sum of pair interaction 
potentials in the Jacobi coordinates in the form:
\begin{equation}
\label{f30aa}
\Delta V(\boldsymbol\rho,\boldsymbol\lambda,\boldsymbol\sigma)=
\Delta V^C(\boldsymbol\rho,\boldsymbol\lambda,\boldsymbol\sigma)+
\Delta V^{conf}(\boldsymbol\rho,\boldsymbol\lambda,\boldsymbol\sigma),
\end{equation}
\begin{equation}
\label{f30b}
\Delta V^C(\boldsymbol\rho,\boldsymbol\lambda,\boldsymbol\sigma)=
\Bigl[
-\frac{2\alpha_{s12}}{3\rho}-\frac{2\alpha_{s34}}{3\lambda}-
\frac{4\alpha_{s13}}{3}\frac{1}{|\boldsymbol\sigma+\frac{m_4}{m_{34}}
\boldsymbol\lambda-\frac{m_2}{m_{12}}\boldsymbol\rho|}-
\frac{4\alpha_{s14}}{3}\frac{1}{|\boldsymbol\sigma-\frac{m_3}{m_{34}}
\boldsymbol\lambda-\frac{m_2}{m_{12}}\boldsymbol\rho|}-
\end{equation}
\begin{displaymath}
\frac{4\alpha_{s23}}{3}\frac{1}{|\boldsymbol\sigma+\frac{m_4}{m_{34}}
\boldsymbol\lambda+\frac{m_1}{m_{12}}\boldsymbol\rho|}-
\frac{4\alpha_{s24}}{3}\frac{1}{|\boldsymbol\sigma-\frac{m_3}{m_{34}}
\boldsymbol\lambda+\frac{m_1}{m_{12}}\boldsymbol\rho|}
\Bigr],
\end{displaymath}
\begin{eqnarray}
\label{f30c}
\Delta V^{conf}(\boldsymbol\rho,\boldsymbol\lambda,\boldsymbol\sigma)=
\Bigl[
\frac{1}{2} A|{\boldsymbol\rho}|+\frac{1}{2} A|{\boldsymbol\lambda}|+
 A|\boldsymbol\sigma+\frac{m_4}{m_{34}}\boldsymbol\lambda-\frac{m_2}{m_{12}}
\boldsymbol\rho|+\\
 A|\boldsymbol\sigma-\frac{m_3}{m_{34}}\boldsymbol\lambda-\frac{m_2}{m_{12}}
\boldsymbol\rho|+
 A|\boldsymbol\sigma+\frac{m_4}{m_{34}}\boldsymbol\lambda+\frac{m_1}{m_{12}}
\boldsymbol\rho|+
 A|\boldsymbol\sigma-\frac{m_3}{m_{34}}
\boldsymbol\lambda+\frac{m_1}{m_{12}}\boldsymbol\rho|+ B
\Bigr], \nonumber
\end{eqnarray}
where $\alpha_{s~ij}$ is the strong interaction constant of the pair of particles $i$ and $j$. 
The parameters in the confinement potential \eqref{f30c} and in the Coulomb potential are chosen 
in exactly the same form as in the case of the interaction of quarks and antiquarks in mesons 
and baryons \cite{tetra2025}. In the Jacobi coordinates, the coordinate wave function has the form:
\begin{equation}
\label{f31}
\Psi(\boldsymbol\rho,\boldsymbol\lambda,\boldsymbol\sigma)=\sum_{I=1}^K
C_I e^{-\frac{1}{2}\bigl[A_{11}(I)
\boldsymbol\rho^2+
2A_{12}(I)\boldsymbol\rho\boldsymbol\lambda+A_{22}(I)\boldsymbol\lambda^2+
2A_{13}(I)\boldsymbol\rho\boldsymbol\sigma+2A_{23}(I)\boldsymbol\lambda\boldsymbol\sigma+
A_{33}(I)\boldsymbol\sigma^2\bigr]},
\end{equation}
where $A_{ij}(I)$ is the matrix of nonlinear variational parameters, $C_I$ are linear variational 
parameters.

The normalization coefficient in wave function of the system in coordinate representation 
is expressed analytically in terms of nonlinear parameters:
\begin{equation}
\label{f32}
{\cal N}=\sum_{I=1}^K\sum_{J=1}^K C_I C_J
\frac{(8\pi^3)^{3/2}}{B_{33}^{3/2}(I,J)(\tilde B_{11}(I,J)\tilde B_{22}(I,J)-
\tilde B_{12}^2(I,J))^{3/2}},
\end{equation}
\begin{equation}
\label{f33}
\tilde B_{11}=B_{11}-\frac{B_{13}^2}{B_{33}},~~~\tilde B_{22}=B_{22}-\frac{B_{23}^2}{B_{33}},~~~
\tilde B_{12}=B_{12}-\frac{B_{13}B_{23}}{B_{33}},~~~B_{kl}=A_{kl}(I)+A_{kl}(J).
\end{equation}

In momentum representation, the wave function of a four-particle system can be obtained 
by performing the Fourier transform \eqref{f31} in the form:
\begin{equation}
\label{f34}
\Psi({\bf p},{\bf q},{\bf t})=\sum_{I=1}^K\frac{C_I}{\sqrt{{\cal N}}(det A)^{3/2}(2\pi)^{\frac{9}{2}}}
e^{-\frac{1}{2det A}[{\bf p}^2(A_{22}A_{33}-
A_{23}^2)+{\bf q}^2(A_{11}A_{33}-A_{13}^2)+{\bf t}^2(A_{11}A_{22}-A_{12}^2)]}
\end{equation}
\begin{displaymath}
e^{-\frac{1}{2det A}[2{\bf p}{\bf t}(A_{12}A_{23}-A_{13}A_{22})+
2{\bf q}{\bf t}(A_{12}A_{13}-A_{11}A_{23})+2{\bf p}{\bf q}(A_{13}A_{23}-A_{12}A_{33})]},
\end{displaymath}
where ${\bf p}$, ${\bf q}$, ${\bf t}$ are relative momenta for particles 12, 34 and the pair 
of particles 12 and 34, and all matrix elements $A_{kl}=A_{kl}(I)$.

The normalization factor of wave function \eqref{f34}, which is obtained after calculating 
the integrals over relative momenta, is equal to 1:
\begin{equation}
\label{f35}
<\Psi|\Psi>=
\int|\Psi({\bf p},{\bf q},{\bf t})|^2\frac{d{\bf p} d{\bf q} d{\bf t}}{(2\pi)^9}=
\sum_{I=1}^K\sum_{J=1}^K \frac{C_I C_J 16\sqrt{2}\pi^{\frac{9}{2}}}
{{\cal N}}
\frac{1}{det Q(I,J)^{\frac{3}{2}}}\frac{1}{det A(I)^{\frac{3}{2}}}
\frac{1}{det A(J)^{\frac{3}{2}}}.
\end{equation}

The elements of the matrix $Q$ in \eqref{f35} are expressed in terms of nonlinear parameters $A_{ij}$ 
as follows:
\begin{eqnarray}
\label{f36}
Q_{11}=\frac{(A_{22}(I)A_{33}(I)-A^2_{23}(I))}{det A(I)}+
\frac{(A_{22}(J)A_{33}(J)-A_{23}(J)^2)}{det A(J)},\\
Q_{22}=\frac{(A_{11}(I)A_{33}(I)-A^2_{13}(I))}{det A(I)}+
\frac{(A_{11}(J)A_{33}(J)-A^2_{13}(J))}{det A(J)},\\
Q_{33}=\frac{(A_{11}(I)A_{22}(I)-A^2_{12}(I))}{det A(I)}+
\frac{(A_{11}(J)A_{22}(J)-A^2_{12}(J))}{det A(J)},\\
Q_{13}=\frac{(A_{12}(I)A_{23}(I)-A_{13}(I)A_{22}(I))}{det A(I)}+
\frac{(A_{12}(J)A_{23}(J)-A_{13}(J)A_{22}(J))}{det A(J)},\\
Q_{23}=\frac{(A_{12}(I)A_{13}(I)-A_{11}(I)A_{23}(I))}{det A(I)}+
\frac{(A_{12}(J)A_{13}(J)-A_{11}(J)A_{23}(J))}{det A(J)},\\
Q_{12}=\frac{(A_{13}(I)A_{23}(I)-A_{12}(I)A_{33}(I))}{det A(I)}+
\frac{(A_{13}(J)A_{23}(J)-A_{12}(J)A_{33}(J))}{det A(J)}.
\end{eqnarray}

Then the magnitude of coordinate wave function at zero relative coordinates, included in 
\eqref{f30}, has the form:
\begin{equation}
\label{f37}
\Psi(0,0,0)=\frac{1}{\sqrt{<\Psi|\Psi>}}\sum_{I=1}^K C_I \frac{1}{16\sqrt{2}\pi^{\frac{9}{2}}}
\frac{1}{det \tilde Q(I)^{\frac{3}{2}}}\frac{1}{det A(I)^{\frac{3}{2}}},
\end{equation}
where matrix elements $\tilde Q$ are determined by first half of terms in matrix elements $Q$.

An increase in calculational accuracy of decay rates can be achieved 
by taking into account various corrections to non-relativistic approximation, including 
relativistic corrections determined by relative momenta in the four-particle system. 
Relativistic corrections in the production of bound states of four particles are determined 
by momentum integrals of the following form:
\begin{equation}
\label{f38}
\langle\frac{{\bf p}^2}{m^2}\rangle=
\int\Psi({\bf p},{\bf q},{\bf k})\frac{{\bf p}^2}{m^2}
\frac{d{\bf p} d{\bf q} d{\bf k}}{(2\pi)^9}=
\sum_{I=1}^K \frac{3C_I}{\sqrt{{\cal N}}}
\frac{(\tilde Q_{22}\tilde Q_{33}-\tilde Q^2_{23})}
{det A(I)^{\frac{3}{2}}det \tilde Q(I)^{\frac{5}{2}}m^2}\times
\end{equation}
\begin{displaymath}
\Bigl[Erf\bigl(\sqrt{\frac{m^2 det\tilde Q}{2(\tilde Q_{22}\tilde Q_{33}-\tilde Q^2_{23}}}\bigr)-
\frac{\sqrt{2}}{3\sqrt{\pi}}\frac{m^3 (det\tilde Q)^{3/2}}{(\tilde Q_{22}\tilde Q_{33}-\tilde Q_{23}^2)^{3/2}}
\bigl(1+\frac{3(\tilde Q_{22}\tilde Q_{33}-\tilde Q_{23}^2)}{m^2 det\tilde Q}\bigr)
e^{-\frac{m^2 det\tilde Q}{2(\tilde Q_{22}\tilde Q_{33}-\tilde Q_{23}^2)}}
\Bigr],
\end{displaymath}
\begin{equation}
\label{f39}
\langle\frac{{\bf q}^2}{m^2}\rangle=
\int\Psi({\bf p},{\bf q},{\bf k})\frac{{\bf q}^2}{m^2}
\frac{d{\bf p} d{\bf q} d{\bf k}}{(2\pi)^9}=
\sum_{I=1}^K \frac{3C_I}{\sqrt{{\cal N}}}
\frac{(\tilde Q_{11}\tilde Q_{33}-\tilde Q^2_{13})}
{det A(I)^{\frac{3}{2}}det \tilde Q(I)^{\frac{5}{2}}m^2}\times
\end{equation}
\begin{displaymath}
\Bigl[Erf\bigl(\sqrt{\frac{m^2 det\tilde Q}{2(\tilde Q_{11}\tilde Q_{33}-\tilde Q^2_{13}}}\bigr)-
\frac{\sqrt{2}}{3\sqrt{\pi}}\frac{m^3 (det\tilde Q)^{3/2}}{(\tilde Q_{11}\tilde Q_{33}-\tilde Q_{13}^2)^{3/2}}
\bigl(1+\frac{3(\tilde Q_{11}\tilde Q_{33}-\tilde Q_{13}^2)}{m^2 det\tilde Q}\bigr)
e^{-\frac{m^2 det\tilde Q}{2(\tilde Q_{11}\tilde Q_{33}-\tilde Q_{13}^2)}}
\Bigr],
\end{displaymath}
\begin{equation}
\label{f40}
\langle\frac{{\bf t}^2}{m^2}\rangle=
\int\Psi({\bf p},{\bf q},{\bf t})\frac{{\bf t}^2}{m^2}
\frac{d{\bf p} d{\bf q} d{\bf t}}{(2\pi)^9}=
\sum_{I=1}^K \frac{3C_I}{\sqrt{{\cal N}}}
\frac{(\tilde Q_{11}\tilde Q_{22}-\tilde Q^2_{12})}
{det A(I)^{\frac{3}{2}}det \tilde Q(I)^{\frac{5}{2}}m^2}\times
\end{equation}
\begin{displaymath}
\Bigl[Erf\bigl(\sqrt{\frac{m^2 det\tilde Q}{2(\tilde Q_{11}\tilde Q_{22}-\tilde Q^2_{12}}}\bigr)-
\frac{\sqrt{2}}{3\sqrt{\pi}}\frac{m^3 (det\tilde Q)^{3/2}}{(\tilde Q_{11}\tilde Q_{22}-\tilde Q_{12}^2)^{3/2}}
\bigl(1+\frac{3(\tilde Q_{11}\tilde Q_{22}-\tilde Q_{12}^2)}{m^2 det\tilde Q}\bigr)
e^{-\frac{m^2 det\tilde Q}{2(\tilde Q_{11}\tilde Q_{22}-\tilde Q_{12}^2)}}
\Bigr],
\end{displaymath}
where Erf is the error function. Despite the convergence of these integrals, when calculating them 
in the region of large values of relative momenta, one integral is cut off at relativistic momentum $m$, 
since in the region of relativistic momenta we have poor knowledge of wave function of the bound 
state.

For numerical calculations of decay width \eqref{f30} with the formation of the tetraquark 
$(cc\bar c\bar c)$, the following values of main parameters are obtained:
\begin{equation}
\label{f41}
\Psi(0,0,0)=0.10~GeV^{9/2}, ~\omega_p=\langle\frac{{\bf p}^2}{m^2}\rangle=0.18,~
\omega_q=\langle\frac{{\bf q}^2}{m^2}\rangle=0.18,~
\omega_t=\langle\frac{{\bf t}^2}{m^2}\rangle=0.12.
\end{equation}

A more detailed calculation of the tetraquark wave functions and parameters is given in \cite{tetra2025}. 
Other relativistic parameters, which are determined by $\langle\frac{{\bf p}{\bf t}}{m^2}\rangle$,
$\langle\frac{{\bf q}{\bf t}}{m^2}\rangle$, $\langle\frac{{\bf p}{\bf q}}{m^2}\rangle$, are omitted here, 
since their values are two orders of magnitude smaller than those considered \eqref{f41}. Numerical 
value of the decay width of $H\to T_{cc\bar c\bar c}+\gamma$ from the $ZZ$ mechanism is presented 
in Table~\ref{tb1}.

\begin{table}[htbp]
\caption{Numerical results for relative decay widths in non-relativistic approximation and 
taking into account relativistic corrections.}
\bigskip
\label{tb1}
\begin{ruledtabular}
\begin{tabular}{|c|c|c|}
Decay & Non-relativistic decay  & Relativistic decay    \\
   &  branching $Br_{nr}$  &  branching $Br_{rel}$   \\       \hline
$H\to ZZ\to T(1^{+-})+\gamma$   &   $ 0.38 \cdot 10^{-14}$  &   $ 0.61 \cdot 10^{-14}$      \\     \hline
$H\to Q\bar Q\to T(1^{+-})+\gamma$   &   $ 0.15 \cdot 10^{-8}$   &    $ 0.21 \cdot 10^{-8}$      \\     \hline
$H\to loop~Q\to T(1^{+-})+\gamma$   &   $ 0.32 \cdot 10^{-11}$   &    $ 0.50 \cdot 10^{-11}$      \\     \hline
$H\to T(1^{--})+\gamma$   &   $ 0.58 \cdot 10^{-10}$   &    $ 0.88 \cdot 10^{-10}$      \\     \hline
\end{tabular}
\end{ruledtabular}
\end{table}

Another important mechanism of tetraquark production is the quark-gluon mechanism, shown in Fig.~\ref{fig1}(a). 
In this case, there are ten amplitudes for the process
$H\to\gamma+T_{cc\bar c\bar c}$, the construction of which is similar to previous amplitudes 
\eqref{f11}-\eqref{f14}. Here, as an example, we give two decay amplitudes in which a photon 
is immediately emitted by a quark or an antiquark produced in the decay of the H-boson:
\begin{equation}
\label{f42}
{\cal M}_{H\to Q\bar Q\to T+\gamma}^{(1)}=\frac{(4\pi\alpha)(4\pi\alpha_s)m}{\sin 2\theta M_Z}
\int d{\bf p}\int d{\bf q}\int d{\bf t}
\bar\Psi_0({\bf p,\bf q,\bf t})D^{\sigma\omega}(k_1)N({\bf p,\bf t})N({\bf q,\bf t})\times
\end{equation}
\begin{displaymath}
Tr \Bigl\{
\hat\Pi_1^V({\bf p,\bf t})
\gamma^\sigma \frac{(\frac{3}{4}\hat T-\hat q+\frac{1}{2}\hat t+m)}{((p_1+p_2+q_2)^2-m^2)}
\frac{(\frac{3}{4}\hat T-\hat r-\hat q+\frac{1}{2}\hat t+m)}{((q_1+k)^2-m^2)}\hat\varepsilon_\gamma
\hat\Pi_2^V({\bf q,\bf t})\gamma^\omega \Bigr\},
\end{displaymath}
\begin{equation}
\label{f43}
{\cal M}_{H\to Q\bar Q\to T+\gamma}^{(2)}=\frac{(4\pi\alpha)(4\pi\alpha_s)m}{\sin 2\theta M_Z}
\int d{\bf p}\int d{\bf q}\int d{\bf t}
\bar\Psi_0({\bf p,\bf q,\bf t})D^{\sigma\omega}(k_1)N({\bf p,\bf t})N({\bf q,\bf t})\times
\end{equation}
\begin{displaymath}
Tr \Bigl\{\hat\Pi_1^V({\bf p,\bf t})
\gamma^\sigma \frac{(\hat r-\hat k-\hat q_1+m)}{((r-k-q_1)^2-m^2)}\hat\varepsilon_\gamma
\frac{(\hat r-\hat q_1+m)}{((r-q_1)^2-m^2)}
\hat\Pi_2^V({\bf q,\bf t})\gamma^\omega \Bigr\}.
\end{displaymath}
The explicit form of remaining decay amplitudes is given in Appendix A.

The color part of amplitudes \eqref{f42}-\eqref{f43} is allocated
as a product of color factors describing pairs of quarks and antiquarks 
in an antisymmetric state:
$\frac{1}{\sqrt{2}}\varepsilon_{ikn}\frac{1}{\sqrt{2}}\varepsilon_{jlm}T^a_{ij}T^a_{kl}$=$-\frac{2}{3}\delta_{nm}$.

Let us simplify the denominators of gluon $D^{\sigma\omega}(k_1)$ and quark propagators in \eqref{f42}-\eqref{f43}
as follows:
\begin{equation}
\label{f44}
\frac{1}{k_1^2}=\frac{4}{M_T^2}\left(1-\frac{4p^2}{M_T^2}-\frac{4q^2}{M_T^2}\right),~~~
\frac{1}{[(r-q_1)^2-m^2]}=\frac{4}{3M_H^2}.
\end{equation}

Such simplifications mean that we neglect the momenta of relative motion of quarks 
in denominators of propagators compared to the Higgs boson or tetraquark masses. 
In the numerator of amplitudes, when transforming them, we retain the second-order terms 
in relative momenta of heavy quarks. As a result, we can factor out the common factor 
from a sum of all ten amplitudes and represent the numerator of total decay amplitude as:
\begin{equation}
\label{f45}
{\cal N_{M}}=\frac{512}{3M_T^2 M_H^2}
\left[10+4\frac{r_3^2}{r_2^2}+\frac{({\bf p}^2+{\bf q}^2)}{m^2}\left(\frac{11}{2}+
4\frac{r_3^2}{r_2^2}\right)+\frac{{\bf t}^2}{m^2}\left(-\frac{1}{12}+\frac{4}{3}\frac{r_3^2}{r_2^2}\right)\right]
\varepsilon_{\mu\nu\lambda\sigma} v^\mu v_\gamma^\nu \varepsilon_T^\lambda \varepsilon_\gamma^\sigma.
\end{equation}

The contribution to the decay width from quark-gluon mechanism with the production 
of a vector tetraquark is then determined by the following expression:
\begin{equation}
\label{f46}
\Gamma(H\to Q\bar Q\to \gamma+T)=\frac{2048\pi^3\alpha^2\alpha_s^2}{9M_Z^8\sin^22\theta_W}
\frac{\left(1-\frac{r_3^2}{r_2^2}\right)^3}{r_3^5 r_2}|\Psi(0,0,0)|^2\times
\end{equation}
\begin{displaymath}
\left[10+4\frac{r_3^2}{r_2^2}+(\omega_p+\omega_q)\left(\frac{11}{2}+
4\frac{r_3^2}{r_2^2}\right)+\omega_t\left(-\frac{1}{12}+\frac{4}{3}\frac{r_3^2}{r_2^2}\right)\right]^2.
\end{displaymath}

Numerical result obtained using \eqref{f46}
is also presented in Table~\ref{tb1}. It follows from the calculations that this decay 
mechanism is dominant. Total decay amplitude is the sum of amplitudes of decays 
$H\to ZZ\to T+\gamma$ and $H\to Q\bar Q\to T+\gamma$. We examine these contributions separately 
to find the leading contribution.
The interference terms are present in total amplitude, but their contribution is suppressed, 
since the contribution of quark-gluon mechanism is dominant. Note that the branching of the decay
$H\to Q\bar Q\to T+\gamma$ is an order of magnitude greater than the branching of the decay
$H\to J/\Psi+J/\Psi$ \cite{apm2023}.

In the study of charmonium pair production in \cite{apm2023} it was shown that it is necessary to consider 
other decay mechanisms, including the quark-gluon loop mechanism. The order of this contribution 
in the interaction constants to the decay width is determined by the factor $\alpha_s^4\alpha^2$. 
In this case, the dependence of the amplitude on the particle mass ratios \eqref{f22} is important \cite{apm2023}. 
Fig.~\ref{fig2} shows two different decay amplitudes that have a quark loop. In the first amplitude, 
a photon is emitted from internal lines of quark loop in Fig.~\ref{fig2}(a), and in the second 
amplitude, a photon is emitted by quarks and antiquarks in the final state. The loop integral 
in the first amplitude is convergent, while the loop integral in the second amplitude diverges 
at large values of loop momentum.

The contribution of first amplitude is determined by two tensor functions, the first of which 
corresponds to the quark loop, and the second to two-gluon production of a tetraquark. 
If a vector tetraquark state is considered, then it arises as a result of vector 
addition of two moments, the spin of two quarks ${\bf S_{cc}}$ and the spin of two antiquarks 
${\bf S_{\bar c\bar c}}$:
\begin{eqnarray}
\label{f46a}
\varepsilon^\mu_{S_{Z~cc}} \varepsilon^\nu_{S_{Z~\bar c\bar c}}<1,1,S_{Z~cc},S_{Z~\bar c\bar c}|0,0>=
\frac{1}{\sqrt{3}}\Bigl(g^{\mu\nu}-v^\mu v^\nu \Bigr),\\
\label{f46b}
\varepsilon^\mu_{S_{Z~cc}} \varepsilon^\nu_{S_{Z~\bar c\bar c}}<1,1,S_{Z~cc},S_{Z~\bar c\bar c}|1,0>=
-\frac{i}{\sqrt{2}}\varepsilon^{\alpha\beta\mu\nu}v_\alpha\varepsilon_{T~\beta},\\
\label{f46c}
\varepsilon^\mu_{S_{Z~cc}} \varepsilon^\nu_{S_{Z~\bar c\bar c}}<1,1,S_{Z~cc},S_{Z~\bar c\bar c}|2,0>=
\varepsilon_T^{\mu\nu}.
\end{eqnarray}
Using \eqref{f46b} in second tensor function, we obtain that the contribution of the amplitude 
in Fig.~\ref{fig2}(a) to the production of a vector tetraquark is equal to 0.

Let us now consider the calculation of a contribution of second interaction amplitude, which 
is shown in Fig.~\ref{fig2}(b). This contribution contains a quark loop, the integral over which 
diverges and can be calculated by usual dispersion method, given that the photon virtualities 
are very different: $k_1^2\sim M_T^2$, $k_2^2\sim M_H^2$. The tensor corresponding to the quark loop 
in Fig.~\ref{fig2}(b) has the following general structure \cite{apm2022}:
\begin{equation}
\label{f47}
T^{\mu\nu}_{Q}=A_{Q}(\tau)(g^{\mu\nu}(v_1v_2)-v_1^\nu v_2^\mu)+
B_{Q}(\tau)[v_2^\mu-v_1^\mu (v_1v_2)][v_1^\nu-v_2^\nu(v_1v_2)],
\end{equation}
where $\tau=\frac{M_H^2}{4m_Q^2}$, $m_Q$ is the mass of the heavy quark in the loop. The structure 
functions $A_{Q}(\tau)$ and $B_{Q}(\tau)$ can be calculated using the explicit expression 
for the loop integral tensor:
\begin{equation}
\label{f48}
A_Q(\tau)=\frac{T^{\mu\nu}}{2}\Bigl[\frac{g^{\mu\nu}}{v_1v_2}-\frac{v_1^\nu v_2^\mu}{(v_1 v_2)^2-v_1^2 v_2^2}\Bigr],
B_Q(\tau)=\frac{T^{\mu\nu}}{2}\Bigl[\frac{3v_1^\nu v_2^\mu}{((v_1 v_2)^2-v_1^2 v_2^2)^2}-
\frac{g^{\mu\nu}}{v_1v_2((v_1 v_2)^2-v_1^2 v_2^2)}\Bigr].
\end{equation}

Since the vertex function \eqref{f3} is proportional to the heavy quark mass $m$, the loop contributions 
of the heavy quarks $b$, $t$ are dominant. Then, at the production of the tetraquark $T_{cc\bar c\bar c}$, 
another mass parameter $M_T/4m_Q\sim m/m_Q$ appears, which can be used for expansions.

Two structure functions $A_Q(\tau)$, $B_Q(\tau)$ can be found using dispersion integrals of the form:
\begin{equation}
\label{f49}
A_Q(\tau)=\frac{1}{\pi}\int_1^\infty\frac{Im A_Q(\tau') d\tau'}{(\tau'-\tau+i0)},~~~
B_Q(\tau)=\frac{1}{\pi}\int_1^\infty\frac{Im B_Q(\tau') d\tau'}{(\tau'-\tau+i0)}.
\end{equation}

The imaginary parts $Im A_Q(\tau)$ and $Im B_Q(\tau)$ can be calculated using the Mandelstam-Cutkosky rule.
Since the main contribution to the Higgs boson decay width is given by the structure function 
$A_Q(\tau)$, we will consider this contribution below. Imaginary part of the function 
$A_Q(\tau)$ is equal to
\begin{equation}
\label{f50}
Im A_Q=\frac{4mr_1^2}{\pi \tau^2}\left[
(\tau-2)\left(
\ln\frac{\sqrt{\tau-1}-\sqrt{\tau}}{\sqrt{\tau-1}+\sqrt{\tau}}-i\pi\right)-2\sqrt{\tau(\tau-1)}
\right],~r_1=\frac{m}{m_Q}.
\end{equation}

By calculating the dispersion integral \eqref{f49}, we obtain the following structure function $A_Q(\tau)$:
\begin{equation}
\label{f51}
A_Q(\tau)=\frac{16mr_1^2}{\pi^2 \tau^2}\left[
(\tau-2)(ArcSin\sqrt{\tau})^2+4\tau-2\sqrt{\tau(1-\tau)}ArcSin\sqrt{\tau}\right].
\end{equation}

Simplifying the denominators of the propagators in four amplitudes in Fig.~\ref{fig2}(b), which 
is carried out similarly to \eqref{f44}, allows us to represent the convolution of the tensor \eqref{f47} 
with total numerator of amplitudes in the form:
\begin{eqnarray}
\label{f52}
{\cal N}=A_{Q}(\tau)\left[g^{\mu\nu}(v_1v_2)-v_1^\nu v_2^\mu\right](N_1^{\mu\nu}+N_2^{\mu\nu}+N_3^{\mu\nu}+N_4^{\mu\nu})=\\
=A_{Q}(\tau)\varepsilon_{\mu\nu\lambda\sigma} v^\mu v_\gamma^\nu \varepsilon^\lambda_T 
\varepsilon^\sigma_T M_T
\left[
8\frac{M_H^2}{M_T^2}+\frac{({\bf p}^2+{\bf q}^2)}{m^2}\left(\frac{4}{3}+\frac{14}{3}\frac{M_H^2}{M_T^2}\right)+
\frac{5}{3}\frac{{\bf t}^2}{m^2}\frac{M_H^2}{M_T^2}\right] \nonumber.
\end{eqnarray}

As a result, the decay width for this quark-gluon loop mechanism will take the form:
\begin{equation}
\label{f53}
\Gamma(H\to loop~Q\to T+\gamma)=\frac{16384\pi m^4\alpha^2\alpha_s^4}{M_Z^{12}\sin 2\theta_W^2 r_2 r_3^9}
\left(1-\frac{r_3^2}{r_2^2}\right)^3\vert\Psi(0,0,0)\vert^2\times
\end{equation}
\begin{displaymath}
\left[8+\frac{14}{3}(\omega_p+\omega_q)+\frac{5}{3}\omega_t\right]^2
\left|\left[(\tau-2)(ArcSin\sqrt{\tau})^2+4\tau-2\sqrt{\tau(1-\tau)}ArcSin\sqrt{\tau}\right]\right|^2.
\end{displaymath}

Table~\ref{tb1} presents total contribution of loops with b and t quarks to the decay width. 
The quark loop mechanism is expected to give a smaller contribution to the decay width compared 
to usual quark-gluon mechanism, which is due to the increase in powers of the constant 
$\alpha_s$. It is useful to consider the production of other vector tetraquarks within 
the quark-gluon mechanism, which is done further for the $1^{--}$ state with orbital angular momentum $L=1$. 
As in the case of previous tetraquark states, we consider states in which $S_{cc}=1$, $S_{\bar c\bar c}=1$, 
but when adding these two spins together, we have a state with total spin $S_T=0$, but with 
total tetraquark angular momentum $J_T=1$. The quantum numbers of such a tetraquark are 
$C_T =(-1)^{S_T+L_T}=(-1)$, $P_T=(-1)^{L_T}=(-1)$. The antisymmetry of wave function 
under the permutation of identical fermions is due to its color part.
To estimate the decay width, we consider such an orbital excitation, which is determined by  
relative orbital momentum $L_\sigma=1$ of one pair $(cc)$ relative to the other pair $(\bar c\bar c)$. 
Appendix B discusses the calculation of such a wave function within the variational method.

General structure of interaction amplitudes is preserved and has the form defined by formulas 
\eqref{f42}, \eqref{f43}, and \eqref{a1}-\eqref{a8} of Appendix A. When adding the spins 
${\bf S}_{cc}$ and ${\bf S}_{\bar c\bar c}$, the projection operator \eqref{f46a} is introduced 
in accordance with the state under consideration. When calculating the interaction amplitudes, 
we take into account linear terms in relative momentum $t$ so that for states with orbital angular 
momentum $L_\sigma=1$, a characteristic integral arises:
\begin{equation}
\label{f54}
\int\frac{d{\bf p} d{\bf q} d{\bf t}}{(2\pi)^9} t_\mu \Psi^{L_\sigma=1}({\bf p},{\bf q},{\bf t})=
(-i)\varepsilon_\mu(L^\sigma_z)R'_\sigma(0,0,0),
\end{equation}
where $\varepsilon_\mu(L^\sigma_z)$ is the polarization vector of orbital motion, $R'_\sigma(0,0,0)$ 
is the derivative of radial wave function with respect to the coordinate $\sigma$ at zero. 
In the case of the production of P-wave states, $R'_\sigma(0,0,0)$ is a new parameter of the theory, 
which can be found within the framework of variational calculation using \eqref{pril2}.

By keeping in the numerator of total production amplitude the quadratic terms in relative momenta 
${\bf p}$, ${\bf q}$, we can represent the expression in a numerator in the form:
\begin{equation}
\label{f55}
{\cal N}(\gamma+T(1^{--}))=(\varepsilon(L^\sigma_z)\varepsilon_\gamma)\left[
13+\frac{52}{3}\frac{r_2^2}{r_3^2}+13\frac{r_3^2}{r_2^2}+\frac{({\bf p}^2+{\bf q}^2)}{m^2}
\left(\frac{97}{12}+\frac{101}{9}\frac{r_2^2}{r_3^2}+\frac{119}{12}\frac{r_3^2}{r_2^2}\right)\right]+
\end{equation}
\begin{displaymath}
(v\varepsilon_\gamma)(v_\gamma\varepsilon(L^\sigma_z))\left[
-\frac{104}{3}+\frac{80}{3}\frac{r_3^2}{r_2^2}-\frac{({\bf p}^2+{\bf q}^2)}{m^2}\left(\frac{202}{9}-
\frac{112}{9}\frac{r_3^2}{r_2^2}\right)\right].
\end{displaymath}

As a result of all transformations, the decay width of the Higgs boson with the production 
of a tetraquark $1^{--}$ is determined by the following expression:
\begin{equation}
\label{f56}
\Gamma(H\to\gamma+T(1^{--}))=\frac{32\pi^3\alpha^2\alpha_s^2\left(1-\frac{r_3^2}{r_2^2}\right)}{M_Z^8r_3r_2^5
\sin^22\theta_W}|R'_\sigma(0,0,0)|^2\left[3n_2^2-(vv_\gamma)^2n_1^2\right],
\end{equation}
\begin{equation}
\label{f57}
n_1=\left[
-\frac{104}{3}+\frac{80}{3}\frac{r_3^2}{r_2^2}-(\omega_p+\omega_q)\left(\frac{202}{9}-
\frac{112}{9}\frac{r_3^2}{r_2^2}\right)\right],
\end{equation}
\begin{equation}
\label{f58}
n_2=\left[
13+\frac{52}{3}\frac{r_2^2}{r_3^2}+13\frac{r_3^2}{r_2^2}+(\omega_p+\omega_q)
\left(\frac{97}{12}+\frac{101}{9}\frac{r_2^2}{r_3^2}+\frac{119}{12}\frac{r_3^2}{r_2^2}\right)\right],
\end{equation}
where $\omega_p$, $\omega_q$ are relativistic parameters determined by relative momenta 
${\bf p}$, ${\bf q}$. 
 
For numerical calculations of decay width \eqref{f56} with the formation of a tetraquark $ (cc\bar c\bar c) $
with orbital momentum 1, the value of a derivative of wave function at zero is obtained:
$R'(0,0,0)=0.02~GeV^{11/2}$. When obtaining this value, it is important to know the wave function of tetraquark
in the field of relativistic momenta. But since this work uses non-relativistic
tetraquark wave function obtained by solving the Schr\"odinger wave equation, error
determination of $ R'(0,0,0) $ is not less than 30 percent.

\section{Conclusion}

The production of exotic quark bound states (tetraquarks, pentaquarks, etc.) is a rapidly developing 
area in particle physics, starting in 2003, when the Belle collaboration \cite{t1} discovered the first 
candidate to tetraquark state. Over the past two decades, many new states have been discovered, the nature 
and properties of which remain to be studied and understood. Heavy tetraquarks, consisting of $c$ and $b$ 
quarks, have a special place among such states, since simple QCD-based models are believed to exist 
for them, allowing calculation of their observed properties. 
The LHCb collaboration \cite{t2}, and later CMS and ATLAS \cite{cms1,t4a}, discovered 
a broad structure with a mass twice the $J/\Psi$ mass and a narrow $X(6900)$ state, which 
is considered to be a bound state of $(cc\bar c\bar c)$.

Along with the calculation of the tetraquark mass spectrum, it is important to calculate the probabilities 
of tetraquark production and study various production mechanisms. The production of heavy tetraquarks 
has already been studied in proton-proton interaction reactions, electron-positron annihilation, 
and in B-meson decays \cite{t8,t9,t9a,t9b,t9c,t10,t11,t12,t13,t14}.
In this paper, we estimate the width of exclusive decay of the Higgs boson with the formation 
of a tetraquark $T_{cc\bar c\bar c}$. Further search for tetraquarks $(cc\bar c\bar c)$ can be carried 
out by their decays: $T\to gg+c\bar c$. Our calculations are based on the relativistic quark model and 
the variational method, within which relativistic effects connected with relative motion of quarks 
\cite{apm2022,apm2023,apm2024} are taken into account. 
To describe the properties of the tetraquark, we use a potential quark model with a Hamiltonian consisting 
of pair Coulomb interactions of quarks and antiquarks arising from the one-gluon exchange potential, 
as well as pair confinement terms. In the nonrelativistic approximation, these terms are supplemented 
by pair hyperfine interaction potentials. When increasing the accuracy of mass spectrum calculation, 
we also took into account relativistic corrections. 
Unlike a number of other calculations, we do not use the smearing of the Dirac $\delta$-function. 
Unlike string quark 
models, in our calculations the confinement potential is related to relative distances between quarks 
and antiquarks. In addition, important differences from other calculations of the masses of bound states 
of four quarks are due to the choice of specific values of quark masses and constants 
of strong interaction between them. As a result, this model allows us to calculate wave functions 
of tetraquarks using the variational method \cite{tetra2025} and use them to calculate the Higgs boson 
decay widths.

The calculations show that among the studied mechanisms of vector tetraquark $(cc\bar c\bar c)$
production, the main one is the quark-gluon mechanism, which gives 
a decay width value several orders of magnitude larger than others. Further study of various mechanisms 
of the tetraquark production with different quantum numbers is of interest in connection 
with the expansion of research in the Higgs sector \cite{enterria}.
It is useful to compare the results of the production of heavy tetraquarks that were obtained 
in different reactions \cite{t7,brodsky1,t9,t9a,t11}. 
Thus, in the proton-proton interaction, the cross sections of the reaction $p+p\to T_{4c}\to J/\Psi+J/\Psi$ 
at the energy $\sqrt{s}=13~TeV$ were obtained in \cite{t8}. Using this cross section and the collider luminosity
$10^{34}~cm^{-2} c^{-1}$, one can estimate the number of such production events at 
$\sim 6000$.
The integrated production rates for S-wave vector $1^{+-}$ state $T_{cc\bar c\bar c}$ was obtained 
in $p-p$ interaction at $\sqrt{13}~TeV$ in \cite{t9b}. Estimated event yield is $(0.5\div 0.8)\cdot 10^9$
assuming an integrated luminosity of 3000 $fb^{-1}$.
The predicted production of vector tetraquark $1^{+-}$ is suppressed relative to $0^{++}$, $2^{++}$ 
by two orders of magnitude.
The Higgs boson decay branchings obtained in this work also allow one to estimate the number of events connected 
with tetraquark production, if one focuses on the Higgs boson factories (FCC-hh (100 TeV, 30 $ab^{-1}$)), 
where the production of up to $2.8\cdot 10^{10}$ Higgs bosons per year is possible \cite{enterria}. 
Thus, the number of events producing tetraquarks $T_{cc\bar c\bar c}$ may 
be about 60. Currently, it is planned to create new high-energy and luminosity electron-positron colliders, 
such as the International Linear Collider (ILC) \cite{ILC}, Compact Linear Collider (CLIC) \cite{CLIC}, which will provide more accurate information about the interactions of the Higgs boson. Such colliders are expected 
to become the basis for the study of New Physics, exotic multiquark states. Given the planned luminosity at 
ILC ${\cal L}=2~ab^{-1}$, corresponding to $\sqrt{s}=250~GeV$ and the luminosity at CLIC ${\cal L}=1~ab^{-1}$ 
at $\sqrt{s}=380~GeV$, total number of tetraquark production events $T_{cc\bar c\bar c}$ in the photoproduction 
reaction is estimated at approximately $9.37\cdot 10^3$ and $8.43\cdot 10^4$ annually \cite{t9a}.

\acknowledgments
The work of F.A.M. is supported by the Foundation for the Development of Theoretical Physics and 
Mathematics BASIS (grant No. 25-1-4-15-1).

\appendix
\section{Production tetraquark $T_{cc\bar c\bar c}$ amplitudes in quark-gluon mechanism}
\label{app1} 

Within the quark-gluon mechanism, there are ten $H\to Q\bar Q\to \gamma+T$ decay amplitudes, two 
of which are given above in \eqref{f42}, \eqref{f43}. The remaining decay amplitudes are of the form:
\begin{equation}
\label{a1}
{\cal M}_{H\to Q\bar Q\to T+\gamma}^{(3)}=\frac{(4\pi\alpha)(4\pi\alpha_s)m}{\sin 2\theta M_Z}
\int d{\bf p}\int d{\bf q}\int d{\bf t}
\bar\Psi_0({\bf p,\bf q,\bf t})D^{\sigma\omega}(k_1)N({\bf p,\bf t})N({\bf q,\bf t})\times
\end{equation}
\begin{displaymath}
Tr \Bigl\{
\hat\Pi_1^V({\bf p,\bf t})
\gamma^\sigma \frac{(\frac{3}{4}\hat T-\hat q+\hat k+\frac{1}{2}\hat t+m)}{((r-q_1)^2-m^2)}
\hat\Pi_2^V({\bf q,\bf t})\gamma^\omega
\frac{(\hat r-\frac{3}{4}\hat T-\hat p+\frac{1}{2}\hat t+m)}{((p_2+k)^2-m^2)}\hat\varepsilon_\gamma
\Bigr\},
\end{displaymath}
\begin{equation}
\label{a2}
{\cal M}_{H\to Q\bar Q\to T+\gamma}^{(4)}=\frac{(4\pi\alpha)(4\pi\alpha_s)m}{\sin 2\theta M_Z}
\int d{\bf p}\int d{\bf q}\int d{\bf t}
\bar\Psi_0({\bf p,\bf q,\bf t})D^{\sigma\omega}(k_1)N({\bf p,\bf t})N({\bf q,\bf t})\times
\end{equation}
\begin{displaymath}
Tr \Bigl\{\hat\Pi_1^V({\bf p,\bf t})\gamma^\sigma 
\frac{(\frac{3}{4}\hat T-\hat q+\hat k+\frac{1}{2}\hat t+m)}{((r-q_1)^2-m^2)}
\hat\Pi_2^V({\bf q,\bf t})\hat\varepsilon_\gamma
\frac{(\frac{3}{4}\hat T-\hat r+\hat q+\frac{1}{2}\hat t+m)}{((k+q_2)^2-m^2)}
\gamma^\omega \Bigr\},
\end{displaymath}
\begin{equation}
\label{a3}
{\cal M}_{H\to Q\bar Q\to T+\gamma}^{(5)}=\frac{(4\pi\alpha)(4\pi\alpha_s)m}{\sin 2\theta M_Z}
\int d{\bf p}\int d{\bf q}\int d{\bf t}
\bar\Psi_0({\bf p,\bf q,\bf t})D^{\sigma\omega}(k_1)N({\bf p,\bf t})N({\bf q,\bf t})\times
\end{equation}
\begin{displaymath}
Tr \Bigl\{
\hat\Pi_1^V({\bf p,\bf t})\gamma^\sigma 
\frac{(\hat r-\hat k-\hat q_1+m)}{((r-k-q_1)^2-m^2)}\hat\varepsilon_\gamma
\frac{(\frac{3}{4}\hat T-\hat q+\frac{1}{2}\hat t+\hat k+m)}{((r-q_1)^2-m^2)}
\hat\Pi_2^V({\bf q,\bf t})
\gamma^\omega
\Bigr\},
\end{displaymath}
\begin{equation}
\label{a4}
{\cal M}_{H\to Q\bar Q\to T+\gamma}^{(6)}=\frac{(4\pi\alpha)(4\pi\alpha_s)m}{\sin 2\theta M_Z}
\int d{\bf p}\int d{\bf q}\int d{\bf t}
\bar\Psi_0({\bf p,\bf q,\bf t})D^{\sigma\omega}(k_1)N({\bf p,\bf t})N({\bf q,\bf t})\times
\end{equation}
\begin{displaymath}
Tr \Bigl\{\hat\Pi_1^V({\bf p,\bf t})\hat\varepsilon_\gamma
\frac{(\frac{1}{4}\hat T+\hat p+\hat k+\frac{1}{2}\hat t+m)}{((p_1+k)^2-m^2)}
\frac{(-\frac{3}{4}\hat T+\hat p+\frac{1}{2}\hat t+m)}{((p_2+q_1+q_2)^2-m^2)}\gamma_\sigma
\hat\Pi_2^V({\bf q,\bf t})
\gamma^\omega \Bigr\},
\end{displaymath}
\begin{equation}
\label{a5}
{\cal M}_{H\to Q\bar Q\to T+\gamma}^{(7)}=\frac{(4\pi\alpha)(4\pi\alpha_s)m}{\sin 2\theta M_Z}
\int d{\bf p}\int d{\bf q}\int d{\bf t}
\bar\Psi_0({\bf p,\bf q,\bf t})D^{\sigma\omega}(k_1)N({\bf p,\bf t})N({\bf q,\bf t})\times
\end{equation}
\begin{displaymath}
Tr \Bigl\{
\hat\Pi_1^V({\bf p,\bf t})
\frac{(-\frac{3}{4}\hat T-\hat k+\hat p+\frac{1}{2}\hat t+m)}{((p_1-r)^2-m^2)}
\gamma^\sigma 
\frac{(-\frac{1}{4}\hat T-\hat k-\hat q+\frac{1}{2}\hat t+m)}{((q_1+k)^2-m^2)}\hat\varepsilon_\gamma
\hat\Pi_2^V({\bf q,\bf t})\gamma^\omega
\Bigr\},
\end{displaymath}
\begin{equation}
\label{a6}
{\cal M}_{H\to Q\bar Q\to T+\gamma}^{(8)}=\frac{(4\pi\alpha)(4\pi\alpha_s)m}{\sin 2\theta M_Z}
\int d{\bf p}\int d{\bf q}\int d{\bf t}
\bar\Psi_0({\bf p,\bf q,\bf t})D^{\sigma\omega}(k_1)N({\bf p,\bf t})N({\bf q,\bf t})\times
\end{equation}
\begin{displaymath}
Tr \Bigl\{\hat\Pi_1^V({\bf p,\bf t})
\frac{(-\frac{3}{4}\hat T+\hat p-\hat k+\frac{1}{2}\hat t+m)}{((r-p_1)^2-m^2)}\gamma_\sigma
\hat\Pi_2^V({\bf q,\bf t})\gamma^\omega
\frac{\hat r-(\frac{3}{4}\hat T-\hat p+\frac{1}{2}\hat t+m)}{((k+p_2)^2-m^2)}
\hat\varepsilon_\gamma
\Bigr\},
\end{displaymath}
\begin{equation}
\label{a7}
{\cal M}_{H\to Q\bar Q\to T+\gamma}^{(9)}=\frac{(4\pi\alpha)(4\pi\alpha_s)m}{\sin 2\theta M_Z}
\int d{\bf p}\int d{\bf q}\int d{\bf t}
\bar\Psi_0({\bf p,\bf q,\bf t})D^{\sigma\omega}(k_1)N({\bf p,\bf t})N({\bf q,\bf t})\times
\end{equation}
\begin{displaymath}
Tr \Bigl\{
\hat\Pi_1^V({\bf p,\bf t})
\frac{(-\frac{3}{4}\hat T-\hat k+\hat p+\frac{1}{2}\hat t+m)}{((p_1-r)^2-m^2)}
\gamma^\sigma \hat\Pi_2^V({\bf q,\bf t})\hat\varepsilon_\gamma
\frac{(\frac{3}{4}\hat T-\hat r+\hat q+\frac{1}{2}\hat t+m)}{((q_2+k)^2-m^2)}\hat\varepsilon_\gamma
\gamma^\omega
\Bigr\},
\end{displaymath}
\begin{equation}
\label{a8}
{\cal M}_{H\to Q\bar Q\to T+\gamma}^{(10)}=\frac{(4\pi\alpha)(4\pi\alpha_s)m}{\sin 2\theta M_Z}
\int d{\bf p}\int d{\bf q}\int d{\bf t}
\bar\Psi_0({\bf p,\bf q,\bf t})D^{\sigma\omega}(k_1)N({\bf p,\bf t})N({\bf q,\bf t})\times
\end{equation}
\begin{displaymath}
Tr \Bigl\{\hat\Pi_1^V({\bf p,\bf t})
\frac{(-\frac{3}{4}\hat T+\hat p-\hat k+\frac{1}{2}\hat t+m)}{((r-p_1)^2-m^2)}
\hat\varepsilon_\gamma
\frac{-(\frac{3}{4}\hat T+\hat p+\frac{1}{2}\hat t+m)}{((p_1+k-r)^2-m^2)}
\gamma_\sigma
\hat\Pi_2^V({\bf q,\bf t})\gamma^\omega
\Bigr\}.
\end{displaymath}

Total decay amplitude of the Higgs boson from quark-gluon mechanism gives the decay width \eqref{f46}.

\section{Tetraquark $T_{cc\bar c\bar c}$ wave function for states with orbital momentum 
$L=1$}
\label{app2} 

To calculate the decay width of the Higgs boson with the production of a tetraquark in the excited state 
$L=1$, it is necessary to know the corresponding wave function. Various orbital excitations are possible 
in a system of four particles, so that the orbital momentum can be represented as the sum of three terms 
related to the relative coordinates $\boldsymbol\rho$, $\boldsymbol\lambda$, $\boldsymbol\sigma$:
\begin{equation}
\label{pril1}
{\bf L}={\bf L}_{\boldsymbol\rho}+{\bf L}_{\boldsymbol\lambda}+{\bf L}_{\boldsymbol\sigma}.
\end{equation}

We restrict ourselves further to the consideration of orbital excitations in the variable ${\boldsymbol\sigma}$, 
which are connected with the orbital moment of one pair of particles $(cc)$ relative to another pair 
of particles $(\bar c\bar c)$. Therefore, we will look for the wave function of the tetraquark 
with ${L}_{\boldsymbol\sigma}=1$ in tensor form:
\begin{equation}
\label{pril2}
\Psi^{L=1}({\boldsymbol\rho},{\boldsymbol\lambda},{\boldsymbol\sigma})=\sum_{I=1}^K 
C_I({\boldsymbol\sigma}{\boldsymbol\varepsilon})
e^{-\frac{1}{2}\bigl[A_{11}(I)
\boldsymbol\rho^2+
2A_{12}(I)\boldsymbol\rho\boldsymbol\lambda+A_{22}(I)\boldsymbol\lambda^2+
2A_{13}(I)\boldsymbol\rho\boldsymbol\sigma+2A_{23}(I)\boldsymbol\lambda\boldsymbol\sigma+
A_{33}(I)\boldsymbol\sigma^2\bigr]},
\end{equation}
where ${\boldsymbol\varepsilon}$ is the polarization vector of orbital motion.

The normalization condition for the wave function will take the form:
\begin{equation}
\label{pril3}
<\Psi^{L=1}|\Psi^{L=1}>=\sum_{I,J=1}^K C_I C_J\frac{12\sqrt{2}\pi^{7/2}(B_{11}B_{22}-B_{12}^2)}{(\det B)^{5/2}}.
\end{equation}

To find the wave functions on the basis of variational method, we calculate the matrix elements 
of kinetic energy operator, which for a system of four particles consists of three terms:
\begin{equation}
\label{pril4}
\hat T=\frac{{\bf p}^2_\rho}{2\mu_1}+\frac{{\bf p}^2_\lambda}{2\mu_2}+\frac{{\bf p}^2_\sigma}{2\mu_3},
\end{equation}
where the reduced masses of particles $\mu_1$, $\mu_2$, $\mu_3$ are introduced:
\begin{equation}
\label{pril5}
\mu_1=\frac{m_1 m_2}{m_{12}},~~~\mu_2=\frac{m_3 m_4}{m_{34}},~~~\mu_3=\frac{m_{12} m_{34}}{m_{1234}}.
\end{equation}

As a result of analytical calculation, we obtain the following expressions for matrix 
elements of the Laplace operators:
\begin{displaymath}
<\Delta_\rho>=\sum_{I,J=1}^K C_I C_J
\frac{12\sqrt{2}\pi^{7/2}}{(\det B)^{7/2}}
\Bigl\{(A_{11}^I)^2\bigl[3B_{11}B_{22}^2B_{33}-3B_{11}B_{22}B_{23}^2-3B_{12}^2B_{22}B_{33}+
5B_{12}^2B_{23}^2-
\end{displaymath}
\begin{displaymath}
4B_{12}B_{13}B_{22}B_{23}+2B_{13}^2B_{22}^2\bigr]+
A_{11}^IA_{12}^I\bigl[B_{11}B_{22}\bigl(10B_{13}B_{23}-
6B_{12}B_{33}\bigr)-4B_{11}B_{12}B_{23}^2+6B_{12}^3B_{33}-
\end{displaymath}
\begin{displaymath}
2B_{12}^2B_{13}B_{23}-4B_{12}B_{13}^2B_{22}\bigr]+
10A_{11}^I A_{13}^I\bigl[B_{11}B_{22}(B_{12}B_{23}-B_{13}B_{22})-
B_{12}^3B_{23}+B_{12}^2B_{13}B_{22}\bigr]+
\end{displaymath}
\begin{displaymath}
A_{11}^I\bigl[3B_{11}^2B_{22}B_{23}^2+6B_{11}B_{12}^2B_{22}B_{33}-3B_{11}B_{12}^2B_{23}^2-
6B_{11}B_{12}B_{13}B_{22}B_{23}+3B_{11}B_{22}(B_{13}^2B_{22}-
\end{displaymath}
\begin{displaymath}
B_{11}B_{22}B_{33})-3B_{12}^4B_{33}+6B_{12}^3B_{13}B_{23}-3B_{12}^2B_{13}^2B_{22}\bigr]+
(A_{12}^I)^2\bigl[3B_{11}^2B_{22}B_{33}+2B_{11}^2B_{23}^2+
\end{displaymath}
\begin{displaymath}
B_{12}^2(5B_{13}^2-3B_{11}B_{33})-4B_{11}B_{12}B_{13}B_{23}-3B_{11}B_{13}^2B_{22}\bigr]+
10A_{12}^IA_{13}^I\bigl[B_{11}B_{12}^2B_{23}+B_{11}B_{22}(B_{12}B_{13}-
\end{displaymath}
\begin{equation}
\label{pril6}
B_{11}B_{23})-B_{12}^3B_{13}\bigr]+5(A_{13}^I)^2(B_{11}B_{22}-B_{12}^2)^2\Bigr\},
\end{equation}
\begin{displaymath}
<\Delta_\lambda>=\sum_{I,J=1}^K C_I C_J
\frac{12\sqrt{2}\pi^{7/2}}{(\det B)^{7/2}}
\Bigl\{(A_{12}^I)^2\bigl[3B_{11}B_{22}^2B_{33}-3B_{11}B_{22}B_{23}^2-3B_{12}^2B_{22}B_{33}+
5B_{12}^2B_{23}^2-
\end{displaymath}
\begin{displaymath}
4B_{12}B_{13}B_{22}B_{23}+2B_{13}^2B_{22}^2\bigr]+
A_{12}^IA_{22}^I\bigl[-6B_{11}B_{12}B_{22}B_{33}-4B_{11}B_{12}B_{23}^2+10B_{11}B_{13}B_{22}B_{23}-
\end{displaymath}
\begin{displaymath}
4B_{12}B_{13}^2B_{22}-2B_{12}^2B_{13}B_{23}+6B_{12}^3B_{33}\bigr]+
10A_{12}^I A_{23}^I\bigl[ (B_{11}B_{22}-B_{12}^2)(B_{12}B_{23}-B_{13}B_{22})\bigr]+
\end{displaymath}
\begin{displaymath}
(A_{22}^I)^2\bigl[3B_{11}^2B_{22}B_{33}+2B_{11}^2B_{23}^2+
B_{12}^2(5B_{13}^2-3B_{11}B_{33})-4B_{11}B_{12}B_{13}B_{23}-3B_{11}B_{13}^2B_{22}\bigr]+
\end{displaymath}
\begin{displaymath}
10A_{22}^IA_{23}^I\bigl[-B_{11}^2B_{22}B_{23}+B_{11}B_{12}^2B_{23}+
B_{11}B_{12}B_{13}B_{22}-B_{12}^3B_{13}\bigr]+5(A_{23}^I)^2(B_{11}B_{22}-B_{12}^2)^2+
\end{displaymath}
\begin{displaymath}
A_{22}^I\bigr[-3B_{11}^2B_{22}^2B_{33}+3B_{11}^2B_{22}B_{23}^2+6B_{11}B_{12}^2B_{22}B_{33}-3B_{11}B_{12}^2B_{23}^2-
6B_{11}B_{12}B_{13}B_{22}B_{23}-
\end{displaymath}
\begin{equation}
\label{pril7}
3B_{12}^4B_{33}+6B_{12}^3B_{13}B_{23}-3B_{12}^2B_{13}^2B_{22}+3B_{11}B_{13}^2B_{22}^2\bigr]\Bigr\},
\end{equation}
\begin{displaymath}
<\Delta_\sigma>=\sum_{I,J=1}^K C_I C_J
\frac{12\sqrt{2}\pi^{7/2}}{(\det B)^{7/2}}
\Bigl\{(A_{13}^I)^2\bigl[3B_{11}B_{22}(B_{22}B_{33}-B_{23}^2)+B_{12}^2(5B_{23}^2-3B_{22}B_{33})-
\end{displaymath}
\begin{displaymath}
4B_{12}B_{13}B_{22}B_{23}+2B_{13}^2B_{22}^2\bigr]+2A_{13}^IA_{23}^I\bigl[
-B_{12}(3B_{11}B_{22}B_{33}+2B_{11}B_{23}^2+2B_{13}^2B_{22})+3B_{12}^3B_{33}+
\end{displaymath}
\begin{displaymath}
5B_{11}B_{13}B_{22}B_{23}-B_{12}^2B_{13}B_{23}\bigr]+10A_{13}^IA_{33}^I\bigl[
(B_{13}B_{22}-B_{12}B_{23})(B_{12}^2-B_{11}B_{22})\bigr]-
\end{displaymath}
\begin{displaymath}
2A_{13}^I(B_{13}B_{22}-B_{12}B_{23})\bigl[B_{11}(B_{23}^2-B_{22}B_{33})+B_{12}^2B_{33}-2B_{12}B_{13}B_{23}+
B_{13}^2B_{22}\bigr]+
\end{displaymath}
\begin{displaymath}
(A_{23}^I)^2\bigl[B_{12}^2(5B_{13}^2-3B_{11}B_{33})-4B_{11}B_{12}B_{13}B_{23}+
B_{11}(3B_{11}B_{22}B_{33}+2B_{11}B_{23}^2-3B_{13}^2B_{22})\bigr]+
\end{displaymath}
\begin{displaymath}
10A_{23}^I A_{33}^I(B_{11}B_{23}-B_{12}B_{13})(B_{12}^2-B_{11}B_{22})-
2A_{23}^I(B_{12}B_{13}-B_{11}B_{23})\det B+
\end{displaymath}
\begin{equation}
\label{pril8}
5(A_{33}^I)^2(B_{12}^2-B_{11}B_{22})^2+5A_{33}^I(B_{12}^2-B_{11}B_{22})\det B
\Bigl\}.
\end{equation}

The matrix elements of the potential energy of pairwise Coulomb interactions are obtained 
in the form:
\begin{equation}
\label{pril8a}
<\frac{1}{\rho}>=\sum_{I,J=1}^K C_I C_J
\frac{8\pi^3}{(\det B)^2(B_{22}B_{33}-B_{23}^2)^{3/2}}
\Bigl[3B_{11}B_{22}(B_{22}B_{33}-B_{23}^2)-
\end{equation}
\begin{displaymath}
B_{12}^2(3B_{22}B_{33}-2B_{23}^2)+2B_{12}B_{13}B_{22}B_{23}-B_{13}^2B_{22}^2\bigr],
\end{displaymath}
\begin{equation}
\label{pril9}
<\frac{1}{\lambda}>=\sum_{I,J=1}^K C_I C_J
\frac{8\pi^3}{(\det B)^2(B_{22}B_{33}-B_{23}^2)^{3/2}}
\Bigl[3B_{11}B_{22}(B_{11}B_{33}-B_{13}^2)-
\end{equation}
\begin{displaymath}
B_{12}^2(3B_{11}B_{33}-2B_{13}^2)+
2B_{11}B_{12}B_{13}B_{23}-B_{11}^2B_{23}^2\bigr],
\end{displaymath}
\begin{equation}
\label{pril10}
<\frac{1}{|{\bf r}_1-{\bf r}_3|}>=\sum_{I,J=1}^K C_I C_J
\frac{2\pi^3}{(\det B)^2 (F_1F_2-F_3^2)^{3/2}}
\Bigl\{B_{11}^2(8B_{22}^2-8B_{22}B_{23}+3B_{22}B_{33}-B_{23}^2)+
\end{equation}
\begin{displaymath}
B_{12}^2\bigl[2(4B_{23}(B_{11}+B_{13})-8B_{11}B_{22}+B_{13}^2-4B_{13}B_{22}+B_{23}^2)-3B_{33}(B_{11}+B_{22})\bigr]+
\end{displaymath}
\begin{displaymath}
2B_{12}(B_{13}-B_{23})\bigl[B_{11}(4B_{22}+B_{23})-B_{13}B_{22}\bigr]+6B_{11}B_{12}B_{22}B_{33}+B_{11}B_{22}
(-3B_{13}^2+8B_{13}B_{22}-
\end{displaymath}
\begin{displaymath}
4B_{13}B_{23}+3B_{22}B_{33}-3B_{23}^2)+8B_{12}^4+B_{12}^3(-8B_{13}+8B_{23}-6B_{33})-B_{13}^2B_{22}^2\Bigr\},
\end{displaymath}
where auxiliary functions are introduced:
\begin{equation}
\label{pril11}
F_1=B_{11}+\frac{1}{4}B_{33}+B_{13},
\end{equation}
\begin{equation}
\label{pril12}
F_2=B_{22}+\frac{1}{4}B_{33}-B_{23},
\end{equation}
\begin{equation}
\label{pril13}
F_3=B_{12}+\frac{1}{2}B_{23}-\frac{1}{4}B_{33}-\frac{1}{2}B_{13}.
\end{equation}

The matrix elements $<\frac{1}{|{\bf r}_1-{\bf r}_4|}>$, $<\frac{1}{|{\bf r}_2-{\bf r}_3|}>$,
$<\frac{1}{|{\bf r}_2-{\bf r}_4|}>$ are obtained from \eqref{pril10} using the substitutions:
$B_{12}\to -B_{12}, B_{23}\to -B_{23}$; $B_{12}\to -B_{12}, B_{13}\to -B_{13}$;
$B_{13}\to -B_{13}, B_{23}\to -B_{23}$.

The matrix elements of the potential energy of paired confinement interactions are obtained 
in the form:
\begin{equation}
\label{pril14}
<{\rho}>=\sum_{I,J=1}^K C_I C_J
\frac{16\pi^3}{(\det B)^3\sqrt{(B_{22}B_{33}-B_{23}^2)}}
\Bigl[3B_{11}B_{22}(B_{22}B_{33}-B_{23}^2)-
\end{equation}
\begin{displaymath}
B_{12}^2(3B_{22}B_{33}-4B_{23}^2)-2B_{12}B_{13}B_{22}B_{23}+B_{13}^2B_{22}^2\bigr],
\end{displaymath}
\begin{equation}
\label{pril15}
<{\lambda}>=\sum_{I,J=1}^K C_I C_J
\frac{16\pi^3}{(\det B)^3\sqrt{(B_{11}B_{33}-B_{13}^2)}}
\Bigl[B_{12}^2(4B_{13}^2-3B_{11}B_{33})-
\end{equation}
\begin{displaymath}
2B_{11}B_{12}B_{13}B_{23}+B_{11}^2(3B_{22}B_{33}+B_{23}^2)-3B_{11}B_{22}B_{13}^2\Bigr],
\end{displaymath}
\begin{equation}
\label{pril16}
<|{\bf r}_1-{\bf r}_3|>=\sum_{I,J=1}^K C_I C_J
\frac{4\pi^3}{(\det B)^3\sqrt{(F_1F_2-F_3^2)}}\Bigl\{
B_{11}^2(16B_{22}^2-16B_{22}B_{23}+3B_{22}B_{33}+B_{23}^2)+
\end{equation}
\begin{displaymath}
B_{12}^2\bigl[4(-8B_{11}B_{22}+4B_{11}B_{23}+B_{13}^2+B_{13}(B_{23}-4B_{22})+B_{23}^2)-
3B_{33}(B_{11}+B_{22})\bigr]+
\end{displaymath}
\begin{displaymath}
2B_{12}(B_{13}-B_{23})(8B_{11}B_{22}-B_{11}B_{23}+B_{13}B_{22})+6B_{11}B_{12}B_{22}B_{33}+B_{11}B_{22}
(-3B_{13}^2+16B_{13}B_{22}-
\end{displaymath}
\begin{displaymath}
8B_{13}B_{23}+3B_{22}B_{33}-3B_{23}^2)+16B_{12}^4-2B_{12}^3(8B_{13}-8B_{23}+3B_{33})+B_{13}^2B_{22}^2
\Bigr\}.
\end{displaymath}

The matrix elements $<|{\bf r}_1-{\bf r}_4|>$, $<|{\bf r}_2-{\bf r}_3|>$,
$<|{\bf r}_2-{\bf r}_4|>$ are obtained from \eqref{pril16} using the substitutions:
$B_{12}\to -B_{12}, B_{23}\to -B_{23}$; $B_{12}\to -B_{12}, B_{13}\to -B_{13}$;
$B_{13}\to -B_{13}, B_{23}\to -B_{23}$.

\end{document}